\documentclass[a4paper,10pt,twoside]{cpc-hepnp}

\usepackage{multicol}
\usepackage{graphicx}
\usepackage{booktabs}
\usepackage{amssymb,bm,mathrsfs,bbm,amscd}
\usepackage[tbtags]{amsmath}
\usepackage{lastpage}
\usepackage{CJK}

\begin{document}
\begin{CJK*}{GB}{gbsn}

\fancyhead[c]{\small Submit to Chinese Physics C
} \fancyfoot[C]{\small \thepage}
\footnotetext[0]{}

\title{Angular Reconstruction of a Lead Scintillating-Fiber Sandwiched Electromagnetic Calorimeter\thanks{
Supported by the China National Science Foundation (10805050)}}

\author{
      Zu-Hao LI(李祖豪)$^{1;1}$\email{lizh@ihep.ac.cn}%
\quad Wei-Wei XU(许伟伟)$^{1}$
\quad Ling-Yu WANG(王玲玉)$^{1}$
\quad Cheng ZHANG(张诚)$^{1}$\\
\quad Zhi-Cheng TANG(唐志成)$^{1}$
\quad Qi YAN(严琪)$^{1}$
\quad Min YANG(杨民)$^{1}$
\quad Yu-Sheng LU(吕雨生)$^{1}$\\
\quad Guo-Ming CHEN(陈国明)$^{1}$
\quad He-Sheng CHEN(陈和生)$^{1}$
}
\maketitle

\address{%
$^1$Institute of High Energy Physics, CAS, Beijing 100049, China}

\begin{abstract}
A new method called Neighbor Cell Deposited Energy Ratio~(NCDER)~is proposed to reconstruct incidence position in a single layer for a 3-dimensional imaging electromagnetic calorimeter~(ECAL).This method was applied to reconstruct the ECAL test beam data for the Alpha Magnetic Spectrometer-02~(AMS-02). The results show that this method can achieve an angular resolution of $7.36\pm0.08^{\circ}/\sqrt{E}\oplus0.28\pm0.02^{\circ}$ in the determination of the photons direction, which is much more precise than that obtained with the commonly-adopted Center of Gravity(COG) method ($8.4\pm0.1^{\circ}/\sqrt{E}\oplus0.8\pm0.3^{\circ}$). Furthermore, since it uses only the properties of electromagnetic showers, this new method could also be used for other type of fine grain sampling calorimeters.
\end{abstract}



\begin{multicols}{2}

\section{Introduction}
The Alpha Magnetic Spectrometer-02~(AMS-02)~is a particle physics detector designed to search for antimatter and dark matter as well as to accurately measure cosmic ray spectra in space~\cite{ref1}. It was installed on the International Space Station~(ISS)~on $19^{th}$ May, 2011 and will record data from cosmic rays for 10 to 20 years. The Electromagnetic Calorimeter~(ECAL)~of the AMS-02 is a fine-grained lead scintillating-fiber sampling calorimeter which allows for precise, 3-dimensional imaging of the longitudinal and lateral shower development. The system provides a high electron/hadron discrimination, as well as good energy and angular resolution~\cite{ref2,ref3}. The structure of the ECAL and the test beam setup is briefly described in Section 2.

A new method, named Neighbor Cell Deposited Energy Ratio~(NCDER)~, was used to reconstruct incidence position in each layer and compared to two alternative methods, Lateral Fit~(LF)~and Center Of Gravity~(COG), the latter being the most frequently used. The comparison shows that the NCDER and the LF methods are more precise than the COG method, and the NCDER method is more efficient than  the LF method (section 5).

\section{AMS-02 ECAL and test beam}
 The AMS-02 ECAL is a 3D imaging calorimeter which consists of a pancake composed from 9 super-layer, giving an active area of 648~$\times$~648 $mm^2$ and a thickness of 166.5 mm. Each super-layer is 18.5mm thick and made of 11 grooved, 1 mm thick lead foils interleaved with layers of 1 mm diameter scintillating fibers, glued together with epoxy resin. The detector imaging capability is obtained by stacking super-layers with fibers alternatively parallel to the x-axis~(5 layers) and y-axis~(4 layers)~(Fig.~\ref{fig_ecal}~a). Each super-layer is read out by 36 Photo Multiplier Tubes (PMTs), arranged alternately on the two opposite ends. Fibers are read out, on one end only, by four anode Hamamatsu PMTs. Each anode covers an active area of $9\times9mm^2$, corresponding to 35 fibers, defined as a cell~(Fig.~\ref{fig_ecal}~b), the minimum detection unit, which corresponds about 1 radiation length and 0.5 Moliere Radius (R$_{M}$). In total the ECAL consists of 18 layers, measuring 10 layers in the y direction and 8 layers in the x direction.~\cite{ref2,ref3}.

    \begin{center}
    \includegraphics[width=8cm]{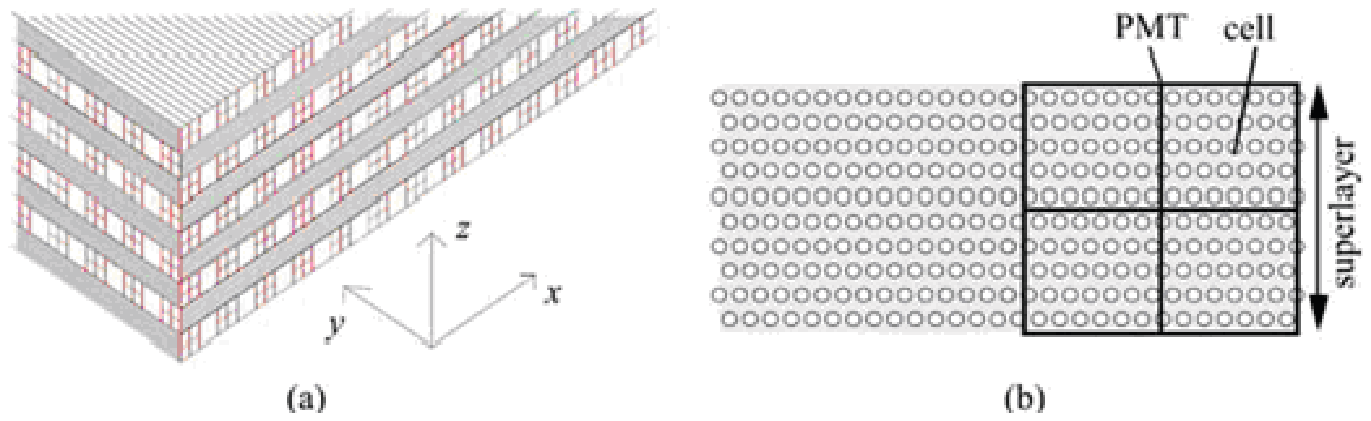}
    \figcaption {\label{fig_ecal}Structure of the AMS-02 ECAL. (a) The superlayer assembly;~(b) the structure of a portion of a superlayer with one PMT.}
    \end{center}

 The flight model of the ECAL was successfully tested and calibrated on the H4 beam line of the Super Proton Synchrotron (SPS) at CERN in July 2007. A schematic diagram of the test beam setup is shown in Fig.~\ref{fig_beamtest}. The ECAL flight model was mounted on a rotating table which can move along the x and y axes and can rotate around the z axis. Events were triggered by four crossed plastic scintillating counters in coincidence. Three silicon tracker ladders were installed in front of the ECAL, which provided accurate information on the beam incident position. Data were taken with a proton beam at 100 GeV and with electron beams at 11 different energies ranging from 6 GeV to 250 GeV. Incidence angles were 0$^{\circ}$, 4.5$^{\circ}$, 7.5$^{\circ}$ and 15$^{\circ}$.~\cite{ref4}.

    \begin{center}
    \includegraphics[width=8cm]{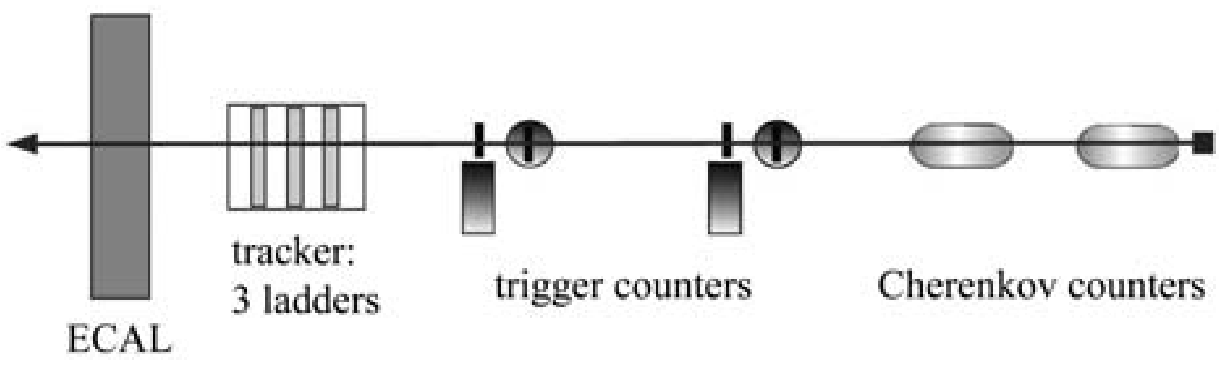}
    \figcaption {\label{fig_beamtest}The AMS02 ECAL test beam setups.}
    \end{center}

\section{Position reconstruction of one single layer}
The most common method for single-layer position reconstruction is the Center of Gravity (COG) method. It takes the center of gravity of the deposited energy as the reconstructed position. The Lateral Fit (LF) method can also be used for position reconstruction. This method uses the two dimensional lateral distribution of deposited energy in a single layer, as described by Equation~\ref{equ_1}~\cite{ref5}. The integral over a cell of the differential energy described by Equation ~\ref{equ_1} is the theoretical value of energy deposited in the cell. Incidence position can be obtained by successfully fitting the theoretical value from measured energy using the Minuit package of ROOT software~\cite{ref6}~.
\begin{equation}\label{equ_1}
\frac{d^2E}{dxdy}=\frac{3\cdot E_{Layer}}{\pi}\cdot \frac{R^2_{Layer}}{(r+R_{Layer})^4}
\end{equation}

The LF method is more precise than the COG method, but due to the complexity of the fit procedure and the limited granularity of the calorimeter, only about 60\% of the events are fitted successfully with this method. Eventually, none of these methods were satisfactory, so we decided to develop a new approach for the reconstruction of electromagnetic showers.

\subsection{Neighbor Cell Deposited Energy Ratio~(NCDER)~Method}

When a high energy electron or photon hit the AMS-02 ECAL, there will be an electromagnetic shower in the ECAL. For a given layer, the cell through which the axis of the shower passes is defined as the central cell. Since the lateral distribution of electromagnetic showers is narrow, the central cell will be the one with most deposited energy.

The energy deposited in the right and left neighbors of the central cell is defined as $E_{Right}$ and $E_{Left}$ respectively. For the AMS-02 ECAL test beam, the incidence position of the particle was obtained precisely from the track reconstruction in the silicon tracker ladders. 

The ratio of $E_{Left}$ to $E_{Right}$ decreases exponentially with the distance of the incidence position to the left edge of the central cell, as shown in Fig.~\ref{fig3}. Fig.~\ref{fig3} is elicited from 100 GeV electron events of AMS-02 test beam data. The dots in the plot represent results from events with 0$^{\circ}$ and 7.5$^{\circ}$ incidence angles, while the line represents the exponential fit curve. Events with 0$^{\circ}$ incidence angle are in perfect match with 7.5$^{\circ}$ events in Fig.~\ref{fig3}, proving that the ratio does not depend on the incidence angle.

    \begin{center}
    \includegraphics[width=7.5cm]{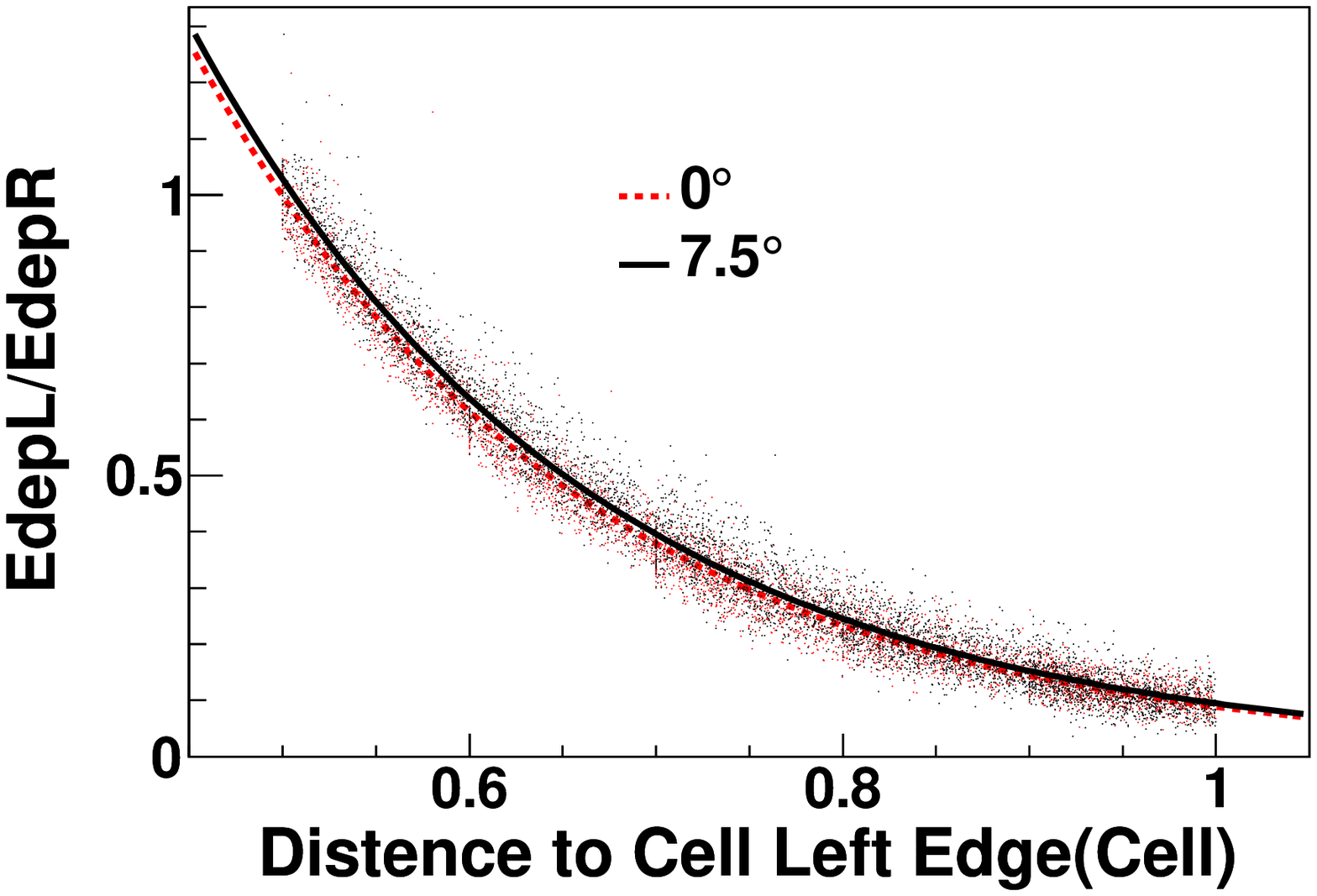}
    \figcaption {\label{fig3}Deposited energy ratio of the left and right neighbor cells vs. the distance from incidence position to the left edge of the cell.}
    \end{center}

The exponential function can be described by two parameters, as Equation~\ref{equ_2}:
\begin{equation}\label{equ_2}
\frac{E_{Left}}{E_{Right}}=e^{A\cdot d +B}
\end{equation}
where $d$ represents the distance from the incidence position to the left edge of the central cell. For electromagnetic showers, when a particle passes across the center of the cell, the deposited energy in the left and right neighbor cells should be equal, thus:
\begin{eqnarray*}
\frac{E_{Left}}{E_{Right}}=e^{A\cdot d +B}=1\\
\Rightarrow 0.5\times A+B=0\\
\Rightarrow B=-0.5\times A
\end{eqnarray*}
So, Equation~\ref{equ_2} can be reduced to Equation~\ref{equ_3}:
\begin{equation}\label{equ_3}
\frac{E_{Left}}{E_{Right}}=e^{\alpha \cdot d -0.5\cdot \alpha}
\end{equation}
The distance from the incidence position to the left edge of the cell is derived from Equation~\ref{equ_4}.
\begin{equation}\label{equ_4}
d=0.5+(ln (\frac{E_{Left}}{E_{Right}}))/\alpha
\end{equation}
Thus, the incidence position $x_{incidence}$ is obtained from Equation~\ref{equ_5}:
\begin{equation}\label{equ_5}
%\begin{split}
x_{incidence}=d+x_{left\_ edge}\\
=0.5+ln(\frac{E_{Left}}{E_{Right}})/\alpha+x_{left\_ edge}
%\end{split}
\end{equation}
$x_{left\_ edge}$ in Equation~\ref{equ_5} represents the position of the left edge of the central cell, which can be obtained from the geometry specifications of the AMS-02 ECAL.
 
 The value of $\alpha$ can be obtained layer by layer by fitting plots simillar to Fig.~\ref{fig3}, but for facility, the dependence of $\alpha$ value on layer number is studied with test beam data. Fig. ~\ref{fig4}, shows the plot of the value of $\alpha$ vs. layer number from the fourth layer onward for 100 GeV electrons. It is shown from the plot that the value of $\alpha$ decreases linearly with the layer number starting from the fourth layer. The value of $\alpha$ can be described as Equation~\ref{equ_6}, where $N_{layer}$ is the layer number, and $P1$ and $P2$ are parameters.
\begin{equation}\label{equ_6}
\alpha=P1\cdot N_{Layer}+P2
\end{equation}

     \begin{center}
     \includegraphics[width=7.5cm]{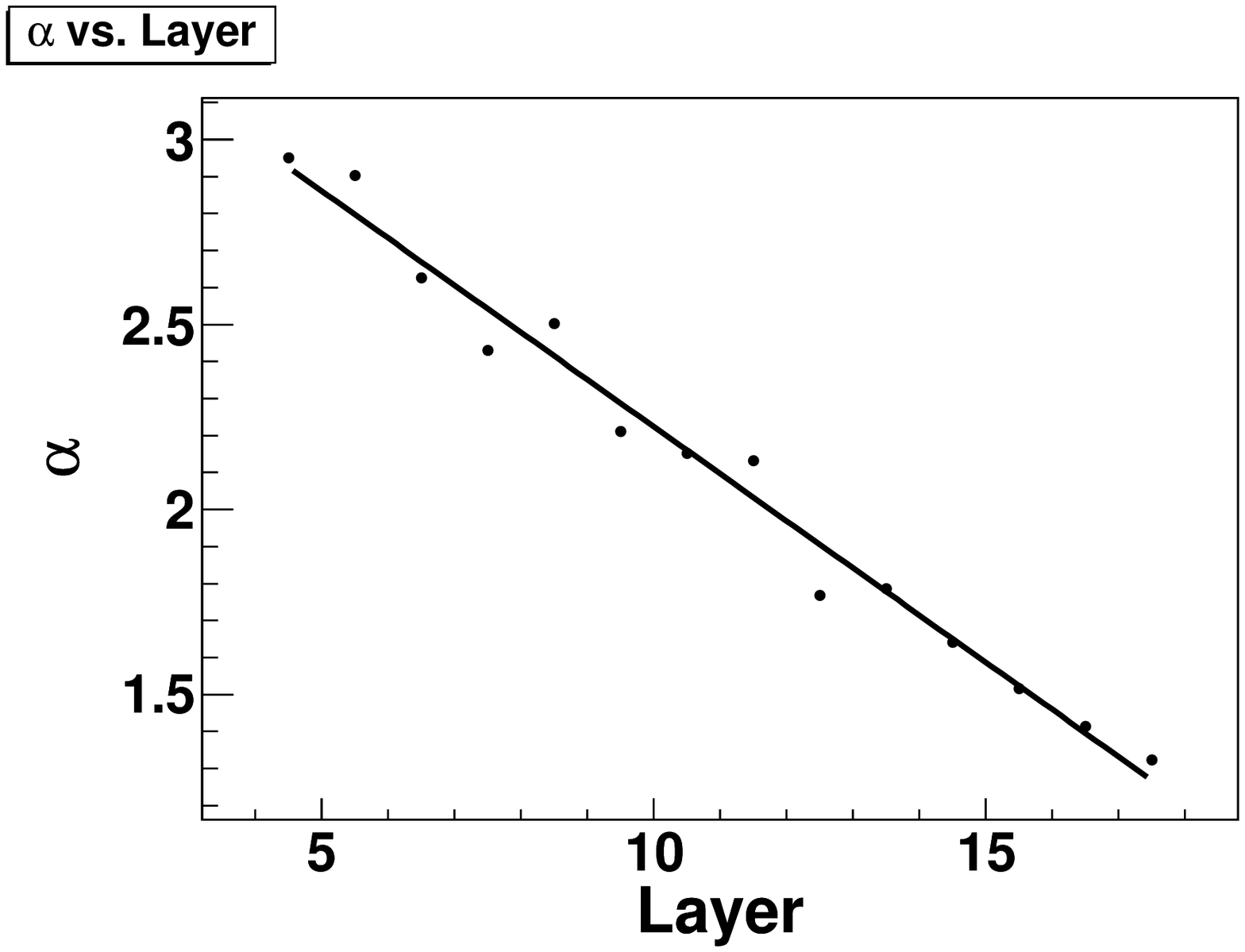}
     \figcaption {\label{fig4}$\alpha$ value vs. layer number.}
     \end{center}

For the first four layers, the electromagnetic shower is not well developed thus the weights for the final direction fitting~\cite{ref8} are very small, so Equation~\ref{equ_6} is used for deriving the value of $\alpha$ for the first three layers.

For 100 GeV electrons, with values of $P1$ and $P2$ taken from the fit in Fig.~\ref{fig4}, $\alpha$ values for all layers can be derived from Equation~\ref{equ_6}, then the incidence position $x_{incidence}$ for all layers can be obtained from Equation~\ref{equ_5}.

For electrons with other energies which were tested in the test beam, $P1$ and $P2$ can be obtained by a similar process. To make this method usable for all energies, $P1$ and $P2$ dependence on energy is studied with test beam data. $P1$ and $P2$ versus energy are shown in Fig. ~\ref{fig5},where stars represent the results obtained from test beam data, while the solid lines are fitting results with empirical functions.

\end{multicols}
%\ruleup
     \begin{center}
     \includegraphics[width=15cm]{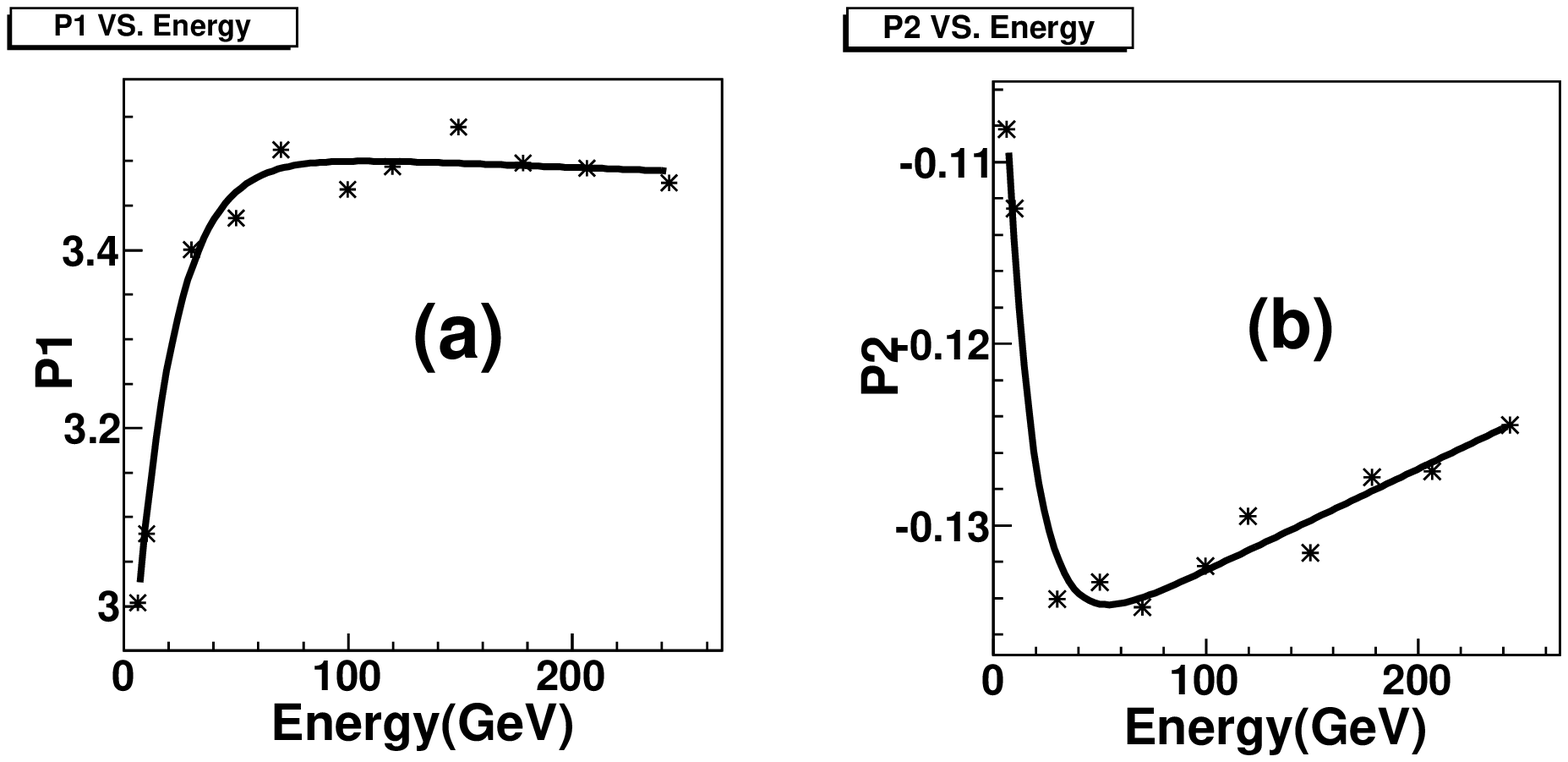}
     \figcaption {\label{fig5}(a)  P1 VS. energy; (b) P2 VS. energy.}
     \end{center}
%\ruledown
\begin{multicols}{2}

For a real event, a preliminary value for the total energy can be reconstructed~\cite{ref7}, then the value of $P1$ and $P2$ can be derived from the empirical functions obtained from Fig. ~\ref{fig5}. Thus, $\alpha$ values for all layers can be obtained from Equation ~\ref{equ_6}, then the incidence position $x_{incidence}$ in all layers can be reconstructed from the $\alpha$ values and energy deposited in cells, using Equation ~\ref{equ_5}.

\subsection{Comparison of the COG, LT and NCDER methods}

A comparison between the incidence position evaluated in the ECAL using the various reconstruction methods, and the position given by the silicon tracker for 100 GeV electrons, are shown in Fig.~\ref{fig6}~(a), (b) and (c). From the figures it is evident that the reconstructed positions with LF and NCDER methods are linear with the position given by the silicon tracker, while that from the COG method is not - it tends to deviate towards the center of the cell.

The distribution of the differences between positions given by silicon trackers and positions reconstructed with these three methods, for 250 GeV electrons with 7.5$^{\circ}$ incidence angle, is plotted in Fig.~\ref{fig6}~(d), which shows that the results with NCDER and LF methods are clearly better than those from the COG method.

\end{multicols}
%\ruleup
    \begin{center}
    \includegraphics[width=15cm]{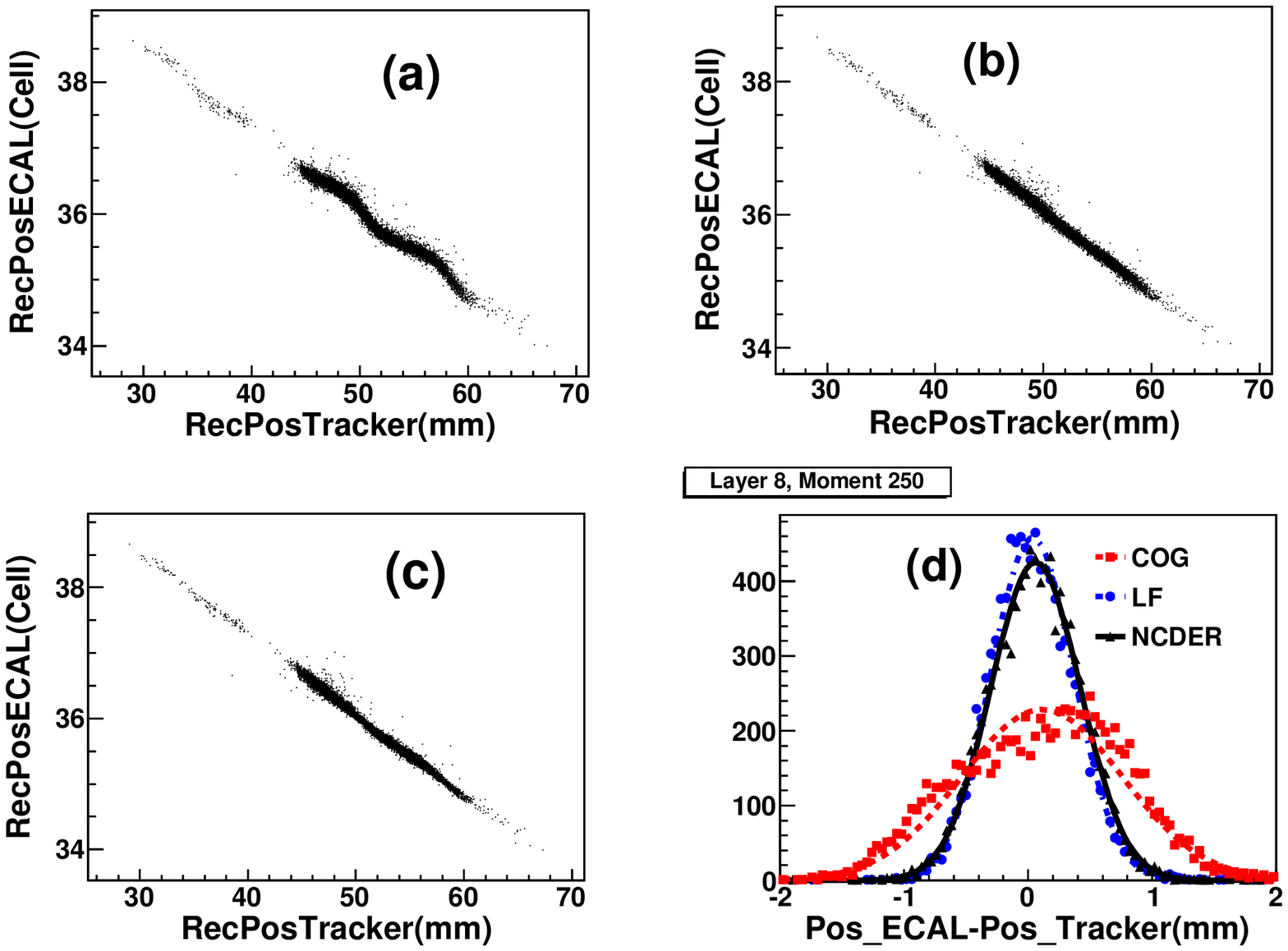}
    \figcaption {\label{fig6} Reconstructed positions in the eighth layer with~
                         (a) the COG method, (b) the NCDER method, (c) the LF method vs. positions measured by the silicon trackers for 100 GeV electrons;
                         (d) distribution of $position\_ rec\_ ECAL - position\_ measure\_ Tracker$ for electrons of 250 Gev with 7.5$^{\circ}$ incidence angle, point line reconstructed using the COG methods, solid and dashed lines for the NCDER and LF methods respectively.}
    \end{center}
%\ruledown
\begin{multicols}{2}

\section{Reconstruction of incidence direction}

 The projection of incidence direction in the x-z plane, which is called $K_{x}$, can be obtained by weighted linear fitting of the reconstructed incidence positions in all 8 layers in the x view, and simillarly for all 10 layers in the y view to obtain the projection in the y-z plane, $K_{y}$.
  
The distribution of $K_{x}$ and $K_{y}$, fitted with positions reconstructed using the COG, NCDER and LF methods for 30 GeV electrons, is shown in Fig.~\ref{fig7}. The beam incidence angles of electrons is 7.5$^{\circ}$ in the y-z projection and 0 in the x-z projection. From the plots, it is clear that the results given by the COG method have the worst resolution, and there is deviation in the K value for the results for inclined events.

\end{multicols}
\begin{center}
\includegraphics[width=12.5cm]{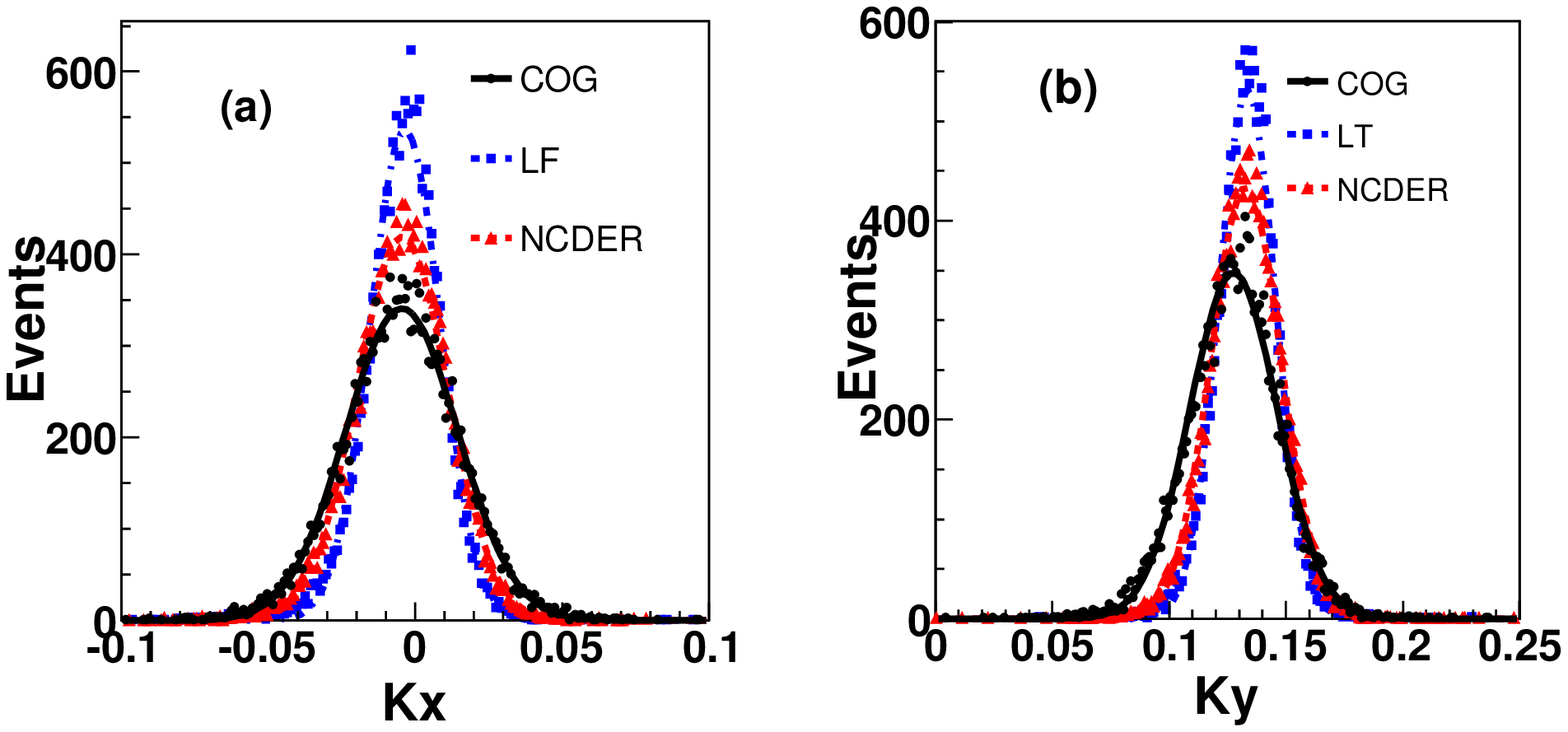}
\figcaption {\label{fig7}$K_{x}$, $K_{y}$ distribution. The circles with solid line are for the COG method, triangles with point-dash line are for the NCDER method, and squares with dashed line are for the LF method.}
\end{center}

\begin{multicols}{2}
The incidence angle can be calculated from $~K_{x}~$and$~K_{y}~$ via the function:~$\theta  = \arctan(\sqrt{K_{x}^2+K_{y}^2})$ and $\Delta \theta = \arctan(\sqrt{(K_{x}-\overline{K_{x}})^2+(K_{y}-\overline{K_{y}})^2}) $. If events with $\Delta \theta$ less than $\theta_{68}$ account for 68$\%$ of the total, $\theta_{68}$ is defined as the angular resolution.

\section{Results and conclusion}

Perpendicular electron events with energy of 6, 10, 30, 50, 70, 100, 120, 150, 180, 200, and 250 GeV were reconstructed with the COF, LF and NCDER methods. Reconstruction with the NCDER method follows the process described above using the same set of parameters. Plots of angular resolution versus energy are shown in Fig.~\ref{fig8}. It shows that angular resolution for results with the NCDER method and the LF method are much better than that from the COG method; for 100GeV electrons, angular resolution improved from 1.10$^{\circ}$ to 0.79$^{\circ}$ and 0.63$^{\circ}$ respectively.

Electron events with energy of 10, 30, 100, 150, 250 GeV and 7.5$^{\circ}$ and 15$^{\circ}$ incidence angles were also reconstructed with the COF, LF and NCDER methods to check the reconstruction quality for inclined events. Reconstructed angles and angular resolutions with the three methods are listed in Tab. \ref{tab1} and Tab. \ref{tab2}. Tab. \ref{tab1} shows that incidence angles reconstructed using the COG method are clearly smaller for electrons with big angles, while ones reconstructed using the other two methods are equal to real values within the system error (The system error for the test beam is 0.2$^{\circ}$; 0.1$^{\circ}$ for precision of rotating table and 0.1$^{\circ}$ for beam orientation.). Table \ref{tab2} shows that angular resolutions change with the energy of the incidence particle, but remain the same for different incidence angles.
    \begin{center}
    \includegraphics[width=7.5cm]{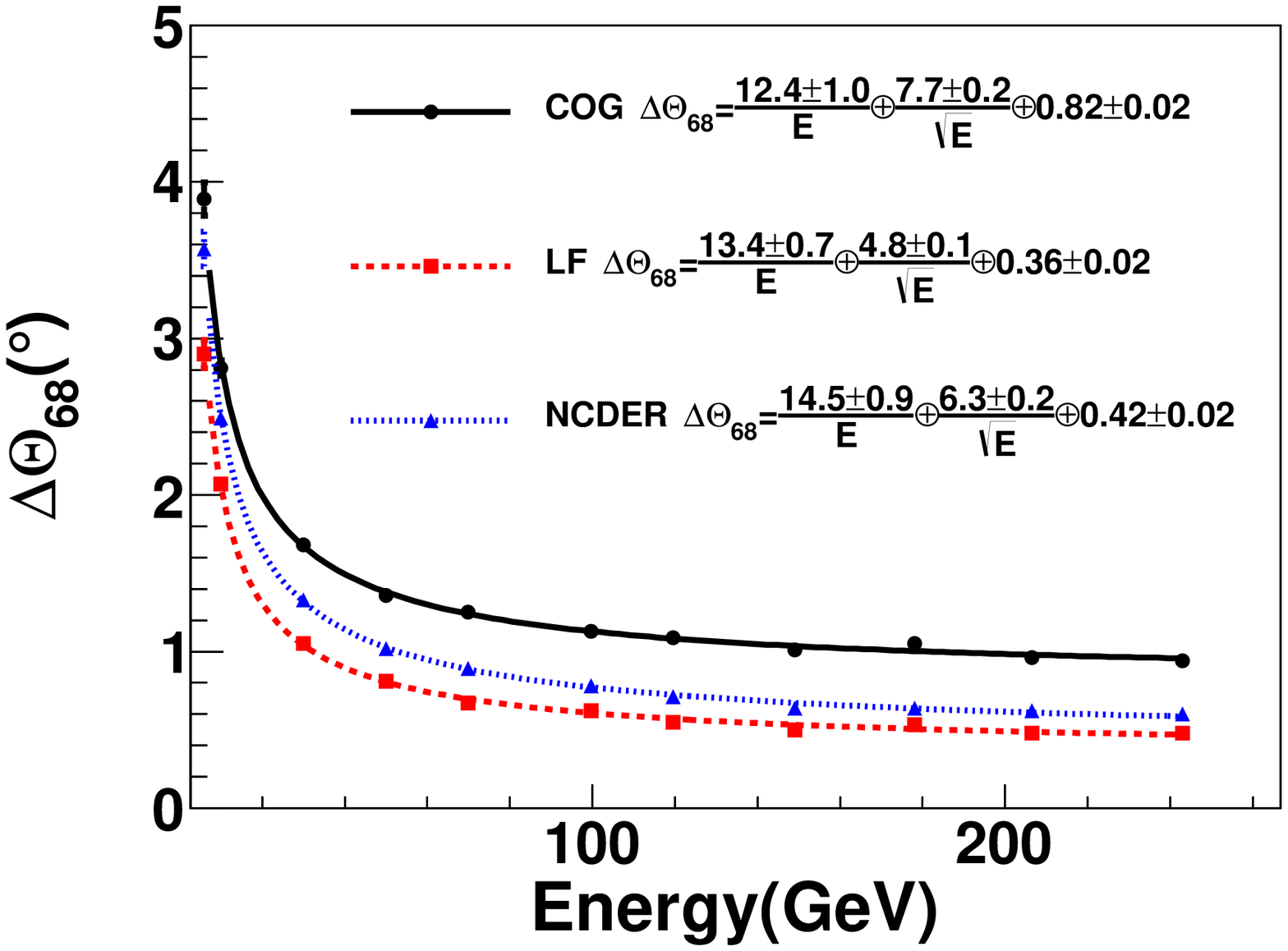}
 \figcaption {\label{fig8}Angular resolution vs. energy. The circles with solid line are for the COG method, triangles with point line are for the NCDER method, and squares with dashed line are for the LF method.}
 \end{center}

\end{multicols}

\vspace{2mm}
  \begin{center}
  \tabcaption{\label{tab1} Reconstructed angles}
\vspace{2mm}
\footnotesize
\begin{tabular*}{170mm}{@{\extracolsep{\fill}}cccccccccccc}
%  \begin{tabular}{cccccccccccc}
\toprule
$Energy(GeV)$ & 10 & 10 & 30 & 30 & 100 & 100 & 100 & 150 & 150 & 250 & 250 \\
\hline
$\theta_{incidence}(^\circ)$ & 7.5 & 15 & 7.5 & 15 &4.5 &7.5 & 15 & 7.5 & 15 &7.5 &15\\
$\theta_{COG}(^\circ)$ &7.4 &14.2 &7.3 &14.1 &4.5 &7.3&14.3 &7.3 &14.3 &7.3 &14.2 \\
$\theta_{NCDER}(^\circ)$ &7.7 &14.8    &7.6 &14.8 &4.7 &7.6&14.8 &7.6 &14.9 &7.5 &14.8\\
$\theta_{LF}(^\circ)$ &7.6 &14.7 &7.6 &14.8&4.7&7.6&14.8 &7.6 &14.9 &7.5 &14.9\\
\bottomrule
\end{tabular*}%
%   \end{tabular}
   \end{center}

   \begin{center}
    \tabcaption{\label{tab2} Angular resolution}
\vspace{2mm}
\footnotesize
\begin{tabular*}{170mm}{@{\extracolsep{\fill}}cccccccccccc}
%    \begin{tabular}{cccccccccccc}
\toprule
$Energy(GeV)$ & 10 & 10 & 30 & 30 & 100 & 100 & 100 & 150 & 150 & 250 & 250 \\
\hline
$\theta_{incidence}(^\circ)$ & 7.5 & 15 & 7.5 & 15 &4.5 &7.5 & 15 & 7.5 & 15 &7.5 &15\\
$\theta_{68COG}(^\circ)$ &2.82 &2.95 &1.64 &1.70 &1.09 &1.04&1.10 &1.02 &1.04 &0.93 &0.84 \\
$\theta_{68NCDER}(^\circ)$ &2.55 &2.72  &1.33 &1.36 &0.80 &0.74&0.76 &0.68 &0.74 &0.58 &0.58\\
$\theta_{68LF}(^\circ)$ &2.10 &2.24 &1.05 &1.10&0.63&0.59&0.62&0.54 &0.59 &0.48 &0.46\\
\bottomrule
   \end{tabular*}
   \end{center}
\begin{multicols}{2}

In conclusion, the NCDER and LF methods are more accurate than the COG method for angular reconstruction. In addition, the reconstructed angle obtained using the COG method is smaller than the real value for inclined tracks, while there is no bias using the LF and NCDER methods. The angular resolution obtained using the LF method is about 20\% better than the one obtained using the NCDER method, but the LF method needs a lot of CPU time and only 60\% of the events can be successfully reconstructed. In comparison, with the NCDER method, almost 100\% of the events are reconstructed, consuming barely any CPU time. Loss of 40\% of the events means lower statistics for the experiment and this could generate a bias in results, which is not acceptable for an experiment like AMS. The optimal method for angular reconstruction is therefor the NCDER method.

\end{multicols}
\centerline{\rule{80mm}{0.1pt}}
\begin{multicols}{2}

\end{multicols}
\clearpage
\end{CJK*}


\begin{thebibliography}{90}
%Section 1
\bibitem{ref1} Battiston R et al. Astropart. Phys., 2000, 13: 51-74
\bibitem{ref2} Antonelli M et al. Nuclear Physics B, 1977, 54: 14-19
\bibitem{ref3} Cadoux F et al. Nuclear Physics B, 2002, 113(Proc.Suppl.):159-165
\bibitem{ref4} Di Falco Stefano. Advances in Space Research, 2010, 45(1): 112-122
\bibitem{ref5} Tao Jun-Quan et al. Chinese Physics C~(HEP \& NP), Vol. 32, No. 3, Mar.,2008: 191-195
\bibitem{ref6}  http://www.root.cern.ch
\bibitem{ref7}  LI Zu-Hao et al. HEP \& NP, 2004, 28: 1182-1188 (in Chinese)
%Section 2
\bibitem{ref8} WANG Xiao-Bin et al. HEP\& NP, 2005, 29: 1071 (in Chinese)

\end{thebibliography}
\end{document}